\documentclass[onecolumn, ,amssym
b, amsmath, nofootinbib,tightenlines, nobibnotes, aps,
prl,epsfig]{revtex4}
 \usepackage{amssymb}
   \usepackage{amsmath}
    \usepackage{amsfonts}
     \usepackage{epsfig}
      \usepackage{bm}

\usepackage{graphicx}
\usepackage{dcolumn}
\usepackage{bm}
\begin{document}
\preprint{APS/123-QED}
\title{Dipole-pion cross section in the saturation regime}

\author{G.R. Boroun}%
 \email{ boroun@razi.ac.ir }
\affiliation{ Physics Department, Razi University, Kermanshah
67149, Iran}

\author{B.Z. Kopeliovich}%
 \email{ boris.kopeliovich@usm.cl}
\affiliation{ Departamento de F\'{i}sica, Universidad T\'{e}cnica Federico Santa Mar\'{i}a, Valparaiso, Chile}

\date{\today}
\begin{abstract}
The scale-dependent dipole-pion cross section is analyzed as a
function of the dipole size $r$ and the impact parameter $b$. This
analysis relies on the DGLAP evolution equation in $\mu \sim 1/r +
\mu_{0}$ at the next-to-leading order (NLO) approximation, with a
specific initial condition at $\mu_{0}$. The dipole-pion cross
section at small Bjorken variable $\beta$ is being considered over
a wide range of transverse separations $\bf r$. Using the Laplace
transformation technique, we describe the determination of the
dipole-pion cross section based on the gluon distribution at the
initial scale $\mu_{0}$ within a kinematic region characterized by
low values of the Bjorken variable $\beta$. We found that
geometric scaling
  for the dipole-pion cross section holds approximately within
  a wide kinematic region of $rQ_s$. The cross section saturates at large dipole
  sizes.\\
\end{abstract}

 \pacs{ 13.60.-r, 13.60.Rj, 13.60.Hb, 14.40.Be}
\keywords{pion, dipole, neutron, crossection} 
\maketitle
\subsection{1. Introduction}

The color-dipole model and dipole cross section were initially
introduced in Ref.\cite{BorisI} to explain that the interaction
eigenstates in Quantum Chromodynamics (QCD) are the eigenstates of the
interaction amplitude with specific values of the transverse
dipole moment. Assuming that gluons in hadrons are concentrated in
small areas occupying only about $10\%$ of the light hadron$^{'}$s total
area can be beneficial in  high-energy hadronic collisions as
mentioned in Ref.\cite{BorisII}. This concept was further developed
beyond the "frozen" dipole approximation in a perturbative manner,
investigating the influence of quantum coherence effects on the
transverse momentum distribution of photons and gluons radiated by
a quark moving through nuclear matter in Ref.\cite{BorisIII}.
Additionally, the nonperturbative interaction for light-cone
fluctuations involving
quarks and gluons is discussed in Ref.\cite{BorisIV}.\\
In the dipole picture \cite{Nikolaev1}, the virtual photon
$\gamma^{*}$ splits into a quark-antiquark pair (a dipole) with
virtuality $Q^2$ exchanged between the electron and target. The
dipole, with the transverse size $\mathbf{r}$ between the quark
and antiquark, interacts with the pion cloud of the proton in the
leading neutrons\footnote{The leading neutrons have been known as
neutron production in deep-inelastic scattering (DIS) on a proton
where neutrons carry a large fraction of the proton$^{,}$s
longitudinal momentum in the forward direction.} \cite{Kumar1,
Kumar2}. The quark and antiquark in this dipole, carry a fraction
$z$ and $1-z$ of the photon$^,$s longitudinal momentum
respectively, probing the pion cloud of the proton in the
inclusive $\gamma^{*}\pi^{*}$ cross section in leading neutron
events $e+p~{\rightarrow}~e'+X+n$. The leading neutron production
in DIS is a method to measure the dipole cross section of the
pion,
$\sigma_{\mathrm{dip}}^{\pi}(x,\mathbf{r})$ \cite{Kopeliovich}.\\
The cross section of the $\gamma^{*}\pi^{*}$ interaction is
directly proportional to the differential cross section for
$\gamma^{*}p{}$ according to the following formula \cite{Carvalho}
\begin{eqnarray}
\sigma^{\gamma^{*}\pi^{*}}(\hat{W}^{2},Q^2)=\frac{1}{f_{\pi/p}(x_{L},t)}
\frac{d^{2}\sigma(W,Q^2,x_{L},t)}{dx_{L}dt},
\end{eqnarray}
where $f_{\pi/p}$ represents the flux of pions emitted by the
proton and explains the splitting of a proton into a $\pi{n}$
system. This flux is well known and can be calculated using chiral
effective theory. In the chiral approach \cite{Thomas}, the proton
is depicted as a combination of states in meson-cloud models
\begin{eqnarray}
|p>&{\rightarrow}&\sqrt{1-a-b}|p_{0}>+\sqrt{a}\bigg{(}-\sqrt{\frac{1}{3}}|p_{0}\pi^{0}>+\sqrt{\frac{2}{3}}|n_{0}\pi^{+}>+...\bigg{)},
\end{eqnarray}
with $a=0.24$ and $b=0.12$. The variables in Eq.(1) are as
follows: $t$ represents the four-momentum transfer squared at the
proton vertex, and $x_{L}$ represents the longitudinal momentum
fraction carried by the outgoing neutron. These variables are
related to the transverse momentum of the neutron, $p_{T}$, as
\begin{eqnarray}
t{\simeq}-\frac{p_{T}^{2}}{x_{L}}-(1-x_{L})\bigg{(}\frac{m_{n}^{2}}{x_{L}}-m_{p}^{2}\bigg{)},
\end{eqnarray}
with the neutron and proton masses (i.e. $m_{n}$ and $m_{n}$).
 The center-of-mass (COM) energies for the photon-proton
and photon-pion systems are denoted by $W$ and $\hat{W}$
respectively, where
$\hat{W}^{2}=(1-x_{L})W^2$.\\
The phenomenon of gluon saturation\footnote{Gluon saturation is a
phenomenon in QCD where, at high energies, the growth of the gluon
density inside a hadron saturates due to the nonlinear
interactions of gluons. } due to the nonlinear effects was
implemented in the dipole picture where has been extended by
authors in Refs.\cite{Amaral, Goncalves, Navarra, Povh}. The color
dipole model in the context of saturation was formulated in
Ref.\cite{Muller}\footnote{In the large-$N_{c}$ approximation, a
Fock component containing gluons can be replaced by a multi-dipole
state.} and later extended by parametrization models incorpo
saturation physics in Refs.\cite{GBW, IP1,IP2, Ferreiro, Kovner}.
Saturation and hadron cross-sections at very high energies
discussed in Ref.\cite{Duraes, Salazar}. At high energies the
small-$x$ gluons in a hadron wavefunction should form a Color
Glass Condensate (CGC) \cite{CGC1, CGC2}, which is characterized
by gluon saturation and the saturation scale $Q_{s}$. This scale
determines the critical line separating the linear and saturation
regimes of QCD dynamics.\\
The total $\gamma^{*}\pi^{*}$ cross section, using the optical
theorem, is related to the dipole-pion cross section as
\begin{eqnarray}
\sigma_{L,T}^{\gamma^{*}\pi^{*}}(\beta,Q^{2})=\int d^{2}\mathbf{r}
\int_{0}^{1} \frac{dz}{4\pi}
|\Psi^{f}_{L,T}(\mathbf{r},z;Q^{2})|^{2}\sigma^{\pi}_{\mathrm{dip}}(\beta,\mathbf{r}),
\end{eqnarray}
where $\Psi_{L,T}(\mathbf{r},z;Q^2)$ are the appropriate spin
averaged light-cone wave functions of the photon, which give the
probability for the occurrence of a $(q\overline{q})$ fluctuation
of transverse size with respect to the photon polarization
\cite{Machado}. The Bjorken variable scaled for the photon-pion
system is given by
\begin{eqnarray}
\beta=\frac{Q^2+m_{f}^{2}}{\hat{W}^{2}+Q^2}=\frac{Q^2+m_{f}^{2}}{(1-x_{L}){W}^{2}+Q^2},
\end{eqnarray}
where $m_{f}$ is the active quark mass defined by the mass of the
charm quark with the number of active flavors
$n_{f}=4$.\\
The dipole-virtual pion cross section was proposed by the GBW
model \cite{GBW} as
\begin{eqnarray}
\sigma^{\pi}_{\mathrm{dip}}(\beta,\mathbf{r})=\sigma_{0}\bigg{(}1-e^{-r^{2}Q^{2}_{s}(\beta)/4}\bigg{)},
\end{eqnarray}
where $Q_{\mathrm{sat}}(\beta)$ plays the role of saturation and
is defined  by the form
$Q_{\mathrm{sat}}^2(\beta)=Q_{0}^{2}(x_{0}/\beta)^\lambda$
 with $Q_{0}^{2}=1~\mathrm{GeV}^2$. The geometric scaling \cite{GS} implies
 that the pion cross section depends only on one dimensionless
 variable $rQ_{s}(\beta)$ (for all values of $r$ and $\beta$), as
 shown by
 \begin{eqnarray}
\sigma^{\pi}_{\mathrm{dip}}(\beta,{r})=\sigma^{\pi}_{\mathrm{dip}}({r}Q_{s}(\beta)).
\end{eqnarray}
The pion cross section can be defined by the following form
\begin{eqnarray}
\sigma^{\pi}_{\mathrm{dip}}(\beta,{r})=\int{d^{2}b}\frac{d\sigma^{\pi}_{\mathrm{dip}}
}{d^{2}b},
\end{eqnarray}
which contains all information about the target and the strong
interaction physics with the impact parameter (IP), $b$.\\

 The
dipole pion cross section at a given impact parameter $b$ (bSat
model or IP-Sat model \cite{IP1, IP2}) contains the DGLAP equation
\cite{DGLAP1, DGLAP2, DGLAP3} for the evolution of the gluon
density at large scales:
\begin{eqnarray}
\frac{d\sigma^{\pi}_{\mathrm{dip}}}{d^{2}b}(\mathbf{b},\mathbf{r},\beta)=2\bigg{[}1-\exp({-\frac{\pi^2r^{2}\alpha_{s}(\mu^2){\beta}g(\beta,\mu^2)T_{\mathrm{\pi}}(\mathbf{b})}{2N_{c}}})\bigg{]},
\end{eqnarray}
with $N_{c}=3$, $\mu^{2}=C/r^{2}+\mu^{2}_{0}$ where $\mu^2$ is the
hard scale, and  the parameters $C$ and $\mu^{2}_{0}$ are obtained
from the fit to the DIS data as summarized in \cite{Zurita,
Kumar2}. The Gaussian form of the function $T_{\pi}(\mathbf{b})$
is determined from the fit to the data as
\begin{eqnarray}
T_{\pi}(\mathbf{b})=\frac{1}{2\pi{B_{\pi}}}\exp{\bigg{(}}-\frac{b^2}{2B_{\pi}}\bigg{)}.
\end{eqnarray}
The parameter $B_{\pi}$ is the width of the pion and is chosen to
be $B_{\pi}=2~ \mathrm{GeV}^{-2}$ from the Belle
measurements \cite{Belle}.\\
Since the free parameters depend on the leading neutron structure
function data or the inclusive proton data, we can consider the
ratio of dipole cross sections at small $x$ given by
\begin{eqnarray}
R_{q}=\frac{\sigma^{\pi}_{\mathrm{dip}}(r,\beta)}{\sigma^{p}_{\mathrm{dip}}(r,\beta)},
\end{eqnarray}
or
\begin{eqnarray}
R_{q}=\frac{\frac{d\sigma^{\pi}_{\mathrm{dip}}}{d^{2}b}(\mathbf{b},\mathbf{r},\beta)}{\frac{d\sigma^{p}_{\mathrm{dip}}}{d^{2}b}(\mathbf{b},\mathbf{r},\beta)}.
\end{eqnarray}
In a constituent quark picture and the color dipole BFKL-Regge
expansion model \cite{Regge}, $R_{q}$ represents the ratio of
valence quarks in the pion and proton i.e., $R_{q}=\frac{2}{3}$.
The value $R_{q}=\frac{1}{2}$ is acceptable as studied in
Ref.\cite{Kopeliovich}.\\
In this paper, we extend the method using a Laplace transform
technique and obtain an analytical method for the solution of the
the dipole-pion cross section in terms of  the known initial
condition in the kinematical region of low values of the Bjorken
variable $\beta$. \\

\subsection{2. Formalism}

The color dipole-pion cross section in the bSat model is given by
\begin{eqnarray}
\frac{d\sigma^{\pi}_{\mathrm{dip}}}{d^{2}b}(\mathbf{b},\mathbf{r},\beta)=2\bigg{[}1-\exp(-F_{\mathrm{DGLAP}}(\mathbf{b},\mathbf{r},\beta))\bigg{]},
\end{eqnarray}
where
\begin{eqnarray}
F_{\mathrm{DGLAP}}(\mathbf{b},\mathbf{r},\beta)=\frac{\pi^2r^{2}\alpha_{s}(\mu^2){\beta}g(\beta,\mu^2)T_{\mathrm{\pi}}(\mathbf{b})}{2N_{c}}.
\end{eqnarray}
The function $F_{\mathrm{DGLAP}}$ is applicable in the DGLAP
evolution equation where the gluon density is dominant at low $x$.
Therefore we find
\begin{eqnarray}
\frac{{\partial}F_{\mathrm{DGLAP}}(\mathbf{b},\mathbf{r},\beta)}{{\partial}r}&=&-\alpha_{s}(\mu^{2})r^2
\frac{{\partial}}{{\partial}r}\bigg{(}\frac{1}{\alpha_{s}(\mu^{2})r^2}
\bigg{)}F_{\mathrm{DGLAP}}(\mathbf{b},\mathbf{r},\beta)-\frac{2C}{r^3\mu^2}{\int_{\beta}^{1}}\frac{\beta}{\gamma^2}d{\gamma}\sum_{n=1}\bigg{(}\frac{\alpha_{s}(\mu^2)}{2\pi}
\bigg{)}^{(n)}\nonumber\\
&&{\times}P^{(n)}_{gg}(\frac{\beta}{\gamma})F_{\mathrm{DGLAP}}(\mathbf{b},\mathbf{r},\gamma),~~~~
\end{eqnarray}
where $n$ represents the order of $\alpha_{s}$. After some
rearranging, we find an evolution equation in terms of $r$
\begin{eqnarray}
{dF_{\mathrm{DGLAP}}(\mathbf{b},\mathbf{r},\beta)}&=&F_{\mathrm{DGLAP}}(\mathbf{b},\mathbf{r},\beta)\bigg{[}\frac{2}{r}+\frac{d{\ln}\alpha_{s}(r)}{dr}\bigg{]}{dr}-\frac{2C}{r^3\bigg{(}\frac{C}{r^2}+\mu_{0}^{2}\bigg{)}}dr{\int_{\beta}^{1}}\frac{\beta}{\gamma^2}d{\gamma}\sum_{n=1}\bigg{(}\frac{\alpha_{s}(\mu^2)}{2\pi}
\bigg{)}^{(n)}\nonumber\\
&&{\times}P^{(n)}_{gg}(\frac{\beta}{\gamma})F_{\mathrm{DGLAP}}(\mathbf{b},\mathbf{r},\gamma).~~~
\end{eqnarray}
We can rewrite the above evolution equation of the dipole-pion
cross section in the bSat model in terms of the variables
$\upsilon=\ln(1/\beta)$ and $r$ instead of $\beta$ and $r$ using
the notation
$\widehat{F}_{\mathrm{DGLAP}}(\mathbf{b},\upsilon,r){\equiv}{F}_{\mathrm{DGLAP}}(\mathbf{b},e^{-\upsilon},r)$
as
\begin{eqnarray}
{d\widehat{F}_{\mathrm{DGLAP}}(\mathbf{b},\upsilon,\mathbf{r})}&=&\widehat{F}_{\mathrm{DGLAP}}(\mathbf{b},\upsilon,\mathbf{r})\bigg{[}\frac{2}{r}+\frac{d{\ln}\alpha_{s}(r)}{dr}\bigg{]}{dr}
-\frac{2C}{r^3\bigg{(}\frac{C}{r^2}+\mu_{0}^{2}\bigg{)}}dr{\int_{0}^{\upsilon}}e^{-(\upsilon-w)}\sum_{n=1}\bigg{(}\frac{\alpha_{s}(\mu^2)}{2\pi}
\bigg{)}^{(n)}\nonumber\\
&&{\times}P^{(n)}_{gg}(\upsilon-w)\widehat{F}_{\mathrm{DGLAP}}(\mathbf{b},\mathbf{r},w)dw,~~~~
\end{eqnarray}
By using the Laplace transform method developed in detail in
\cite{Block1, Block2, Block3, Block4, Boroun1} as
$\mathcal{L}[\widehat{F}_{\mathrm{DGLAP}}(\mathbf{b},\upsilon,\mathbf{r});s]{\equiv}{F}_{\mathrm{DGLAP}}(\mathbf{b},s,\mathbf{r})$
and the fact that the Laplace transform of a convolution factors
is simply the ordinary product of the Laplace transform of the
factors, we find
\begin{eqnarray}
F_{\mathrm{DGLAP}}(\mathbf{b},s,{r})=F_{\mathrm{DGLAP}}(\mathbf{b},s,r_{0})\bigg{(}\frac{r}{r_{0}}\bigg{)}^{2}\frac{\alpha_{s}(r)}{\alpha_{s}(r_{0})}
\exp\bigg{[}-\int_{r_{0}}^{r}\frac{2C}{\xi^3\bigg{(}\frac{C}{\xi^2}+\mu_{0}^{2}\bigg{)}}
\sum_{n=1}\bigg{(}\frac{\alpha_{s}(\xi)}{2\pi}
\bigg{)}^{(n)}h^{(n)}_{gg}(s) d{\xi}\bigg{]}.
\end{eqnarray}
To find the inverse Laplace transform of the factors, we find that
\begin{eqnarray}
\widehat{F}_{\mathrm{DGLAP}}(\mathbf{b},\upsilon,r){=}\int_{0}^{\upsilon}
\widehat{\eta}(\mathbf{b},w,r,r_{0})\widehat{J}(\upsilon-w,\alpha_{s}(r))dw,
\end{eqnarray}
where
\begin{eqnarray}
\widehat{\eta}(\mathbf{b},\upsilon,r,r_{0})=\widehat{F}_{\mathrm{DGLAP}}(\mathbf{b},\upsilon,r_{0})\bigg{(}\frac{r}{r_{0}}\bigg{)}^{2}\frac{\alpha_{s}(r)}{\alpha_{s}(r_{0})}.
\end{eqnarray}
The inverse Laplace transform of the coefficients
$h_{gg}^{(n)}(s)$ in Eq. (18) is straightforward, keeping the
$1/s$ terms of the coefficients at the high-energy region as
\begin{eqnarray}
\widehat{J}(\upsilon,\alpha_{s}(r))=\delta(\upsilon)+\frac{\sqrt{-\phi}}{\sqrt{\upsilon}}\mathrm{I_{1}}(2\sqrt{-\phi}\sqrt{\upsilon}),
\end{eqnarray}
where\footnote{$I_{1}(x)$ is the Bessel function.}
\begin{eqnarray}
\phi=\frac{2C_{A}}{2\pi}\int_{r_{0}}^{r}\frac{2C\alpha_{s}(r)}{r^3\bigg{(}\frac{C}{r^2}+\mu_{0}^{2}\bigg{)}}dr+\frac{(12C_{F}T_{f}-46C_{A}T_{f})}{9(2\pi)^{2}}\int_{r_{0}}^{r}\frac{2C\alpha^{2}_{s}(r)}{r^3\bigg{(}\frac{C}{r^2}+\mu_{0}^{2}\bigg{)}}dr,
\end{eqnarray}
with $C_{F}=\frac{N_{c}^{2}-1}{2N_{c}}$, $T_{f}=\frac{1}{2}n_{f}$
and $C_{A}=3$. Transforming back in to $\beta$ space,
${F}_{\mathrm{DGLAP}}(\mathbf{b},\beta,r)$ is given by
\begin{eqnarray}
{F}_{\mathrm{DGLAP}}(\mathbf{b},\beta,r)=\bigg{(}\frac{r}{r_{0}}\bigg{)}^{2}\frac{\alpha_{s}(r)}{\alpha_{s}(r_{0})}\bigg{[}{F}_{\mathrm{DGLAP}}(\mathbf{b},\beta,r_{0})
+\int_{\beta}^{1}{F}_{\mathrm{DGLAP}}(\mathbf{b},\gamma,r_{0})\frac{\sqrt{-\phi}}{\sqrt{{\ln}\frac{\gamma}{\beta}}}\mathrm{I_{1}}(2\sqrt{-\phi}\sqrt{{\ln}\frac{\gamma}{\beta}})\frac{d\gamma}{\gamma}\bigg{]},
\end{eqnarray}
where
\begin{eqnarray}
{F}_{\mathrm{DGLAP}}(\mathbf{b},\beta,r_{0})=\frac{\pi^{2}r_{0}^{2}\alpha_{s}(r_{0}){\beta}g({\beta},r_{0})T_{\pi}(\mathbf{b})}{2N_{c}},
\end{eqnarray}
and the initial gluon distribution at the scale
$\mu_{0}^{2}=1.1~\mathrm{GeV}^2$ is defined in the form
\cite{Kumar2}
\begin{eqnarray}
{\beta}g(\beta,\mu_{0}^{2})=A_{g}\beta^{-\lambda_{g}}(1-\beta)^{6},
\end{eqnarray}
where parameters in the bSat model are motivated by the leading
neutron structure function HERA data for $\beta{\leq}0.01$
\cite{Kumar1, Kumar2}. Therefore, the evolution of the color
dipole-pion cross section in the bSat model is defined from an
exclusive measurement of $e+p~{\rightarrow}~e+J/\Psi+\pi+n$ by the
following form
\begin{eqnarray}
\frac{1}{2}\frac{d\sigma^{\pi}_{\mathrm{dip}}}{d^{2}b}(\mathbf{b},\mathbf{r},\beta)=1-\exp{\bigg{\{}}-\bigg{(}\frac{r}{r_{0}}\bigg{)}^{2}\frac{\alpha_{s}(r)}{\alpha_{s}(r_{0})}\bigg{[}{F}_{\mathrm{DGLAP}}(\mathbf{b},\beta,r_{0})
+\int_{\beta}^{1}{F}_{\mathrm{DGLAP}}(\mathbf{b},\gamma,r_{0})\frac{\sqrt{-\phi}}{\sqrt{{\ln}\frac{\gamma}{\beta}}}\mathrm{I_{1}}(2\sqrt{-\phi}\sqrt{{\ln}\frac{\gamma}{\beta}})\frac{d\gamma}{\gamma}\bigg{]}\bigg{\}}.~~
\end{eqnarray}
The Bjorken scaling in the photon-proton scattering in the dipole
picture is given by
\begin{eqnarray}
x=\frac{(\mu^2+m_{c}^{2})\beta(1-x_{L})}{(1-{\beta}x_{L}){\mu}^{2}+m_{c}^2},
\end{eqnarray}
where
\begin{eqnarray}
\beta=\frac{\mu^2+m_{c}^{2}}{(1-x_{L}){W}^{2}+\mu^2},
\end{eqnarray}
with $m_{c}=1.3~\mathrm{GeV}$ and $x_{L}=0.6$.\\
In this section, we have summarized the Laplace transform method
for obtaining an analytical solution for the dipole-pion cross
section. Starting with the explicit form of the dipole-pion cross
section [i.e., Eq. (26)], we will now extract numerical results
for small $\beta$ across a wide range of $\mu^2$ values. This will
be done using the initial gluon distribution\footnote{In the
Jefferson Lab Angular Momentum (JAM) collaboration, the pion$^,$s
parton distribution functions (PDFs) have been studied within a
Bayesian Monte Carlo framework with threshold re-summation and
transverse momentum ($p_{T}$) at the scales $\mu=\mu_{0}=m_{c}$
and $\mu^2=10~\mathrm{GeV}^2$. These
finding have been reported in Refs.\cite{JAM1, JAM2}.} at the scale $\mu_{0}^{2}$ in the next section.\\

\subsection{3. Results and Conclusion}

The free parameters for the pion and proton are selected from the
fit results in Refs.\cite{Kumar2, Zurita, Golec}, as summarized in
Table I. The QCD parameter $\Lambda$ for four  active flavors has
been determined as $\alpha_{s}(M_{z}^{2})=0.1166$ resulting in
$\Lambda^{\mathrm{LO}}_{nf=4}=136.8~\mathrm{MeV}$ and
$\Lambda^{\mathrm{NLO}}_{nf=4}=284.0~\mathrm{MeV}$. Table I
displays the free parameters for the pion \cite{Kumar2, Zurita}
and proton \cite{ Golec}.
\begin{table}
\centering \caption{Free parameters for pion \cite{Kumar2, Zurita}
and proton \cite{ Golec}.
  }\label{table:table1}
\begin{minipage}{\linewidth}
\renewcommand{\thefootnote}{\thempfootnote}
\centering
\begin{tabular}{|l|c|c|c|c|c|} \hline\noalign{\smallskip}
$---$  & $A_{g}$  & $\lambda_{g}$ & $C$ & $\mu_{0}^{2}~[\mathrm{GeV}^2]$ &  $B~[\mathrm{GeV}^{-2}]$ \\
\hline\noalign{\smallskip}
$\mathrm{Proton}$ &  1.350 & 0.079 & 0.380 & 1.73 & 4.0  \\
$\mathrm{Pion }$ &  1.208 & 0.060 & 1.453 & 1.10 & 2.3   \\
\hline\noalign{\smallskip}
\end{tabular}
\end{minipage}
\end{table}
The dipole-pion cross sections,
$\frac{d\sigma^{\pi}_{\mathrm{dip}}}{2d^{2}b}$, are plotted in
Fig.1 over a wide range of the dipole size $r$ in the bSat model
at different values of the scaled Bjorken variable $\beta=10^{-2}$
and $10^{-4}$ with the impact parameters $b=0$ and
$2~\mathrm{GeV}^{-1}$.\\
In Fig.1 (left-hand side), the dipole-pion cross sections are
shown at $\beta=10^{-2}$ (upper panel) and $\beta=10^{-4}$ (lower
panel) as a function of dipole size $r$. The dipole-pion cross
sections saturate for large dipole sizes and increase towards
lower dipole sizes as the Bjorken scaling decreases. The impact
parameters (i.e., $b=0$ and $2~\mathrm{GeV}^{-1}$) affect the
dipole-pion cross sections at various values of $r$. In Fig.1
(right-hand side), the dipole-pion cross sections in the bSat
model converge into a single curve when plotted against the
dimensionless variable $rQ_{s}$. The saturation
formalism\footnote{An important property of the saturation
formalism is the geometric scaling phenomenon, which means that
the scattering amplitude and corresponding cross sections can
scale on the dimensionless scale $rQ_{s}$.} of the dipole-pion
cross section leads
 to improved results
 because the evolution of the cross
section due to the Laplace transform becomes a function of a
single variable, $rQ_{s}$, for almost all values of $r$ with
different $\beta$ at $b=0$ and $2~\mathrm{GeV}^{-1}$ in the bSat
model.\\
 Indeed, we observe that the presence of geometric scaling
in the GBW model is illustrated in the dipole-pion cross sections.
All the curves from the left plot with different $\beta$ and $b$
values merge into a single curve when the dipole-pion cross
section is plotted against the dimensionless variable $rQ_{s}$.
This shows that the dipole cross section is indeed a function of a
single dimensionless variable and is in line with the GBW model.
Indeed, the dipole pion cross sections resulting from the
evolution method based on the Laplace transform do not break the
geometrical scaling in a wide range of $rQ_{s}$. In
Ref.\cite{Kumar2}, the author demonstrates that the dipole cross
section for the bSat model with different $\beta$ values does not
merge into a single curve when plotted against the dimensionless
variable $rQ_{s}$ due to the breaking of geometric scaling. The
author in Ref.\cite{Kumar2} suggests that this is because of the
explicit DGLAP evolution of the gluon density in the bSat model
where the gluon density is evaluated at a scale $\mu \sim 1/r +
\mu_{0}$. We improve the results by addressing the breaking of
geometrical scaling caused by the
 DGLAP evolution based on the Laplace transform, as shown in the right panel
 of Fig.1.\\

The values of $R_{q}$ predicted from the ratio $
{\frac{d\sigma^{\pi}_{\mathrm{dip}}}{d^{2}b}(\mathbf{b},\mathbf{r},\beta)}/{\frac{d\sigma^{p}_{\mathrm{dip}}}{d^{2}b}(\mathbf{b},\mathbf{r},\beta)}
$ in a wide range of dipole sizes $r$ in the bSat model at
different values of the scaled Bjorken variable $\beta=10^{-2}$
and $10^{-4}$ with impact parameters $b=0$ and
$2~\mathrm{GeV}^{-1}$ are illustrated in Fig.2. This ratio varies
over a wide range of dipole sizes $r$ \cite{Kumar2, Kopeliovich}
and is independent of $\beta$ at large and low values of $r$ (left
panel of Fig.2). It is observed that the saturation scale
increases towards lower values of $r$ as $\beta$ decreases. The
behavior of $R_{q}$ depends on the impact parameter $b$ at
 small values of $r$ and saturates at large values of $r$. The minimum values of $R_{q}$ are
obtained due to the
$\beta$ and $b$ values in Table II.\\
\begin{table}
\centering \caption{Minimum values of $R_{q}$ into $r$
($0.01<r<0.07~\mathrm{fm}$)
  }\label{table:table1}
\begin{minipage}{\linewidth}
\renewcommand{\thefootnote}{\thempfootnote}
\centering
\begin{tabular}{|l|c|c|c|} \hline\noalign{\smallskip}
$\mathrm{Min}$  & $\beta$  & $b=0~\mathrm{GeV}^{-1}$ & $b=2~\mathrm{GeV}^{-1}$  \\
\hline\noalign{\smallskip}
$R_{q}$ &  $10^{-2}$ & $\sim 0.61$ & $\sim 0.42$  \\
$R_{q}$ &  $10^{-4}$ & $\sim 0.68$ & $\sim 0.47$  \\
\hline\noalign{\smallskip}
\end{tabular}
\end{minipage}
\end{table}
In Fig.2 (left-hand side), $R_{q}$ is plotted at $\beta=10^{-2}$
(upper panel) and $\beta=10^{-4}$ (lower panel) as a function of
dipole size $r$.  $R_{q}$ saturates at large dipole sizes, with
saturation increasing towards lower dipole sizes as the Bjorken
scaling decreases. In Fig.2 (right- hand side), we observe that
the behavior of $R_{q}$ is plotted independently of $\beta$ at
both large and small values of $r$ for the impact parameters $b=0$
(upper) and $2~\mathrm{GeV}^{-1}$ (lower). This value depends on
$\beta$ at moderate $r$
($2{\times}10^{-2}{\lesssim}r{\lesssim}7{\times}10^{-1}~\mathrm{fm}$) as the impact parameter increases.\\
\begin{figure}
\includegraphics[width=0.8\textwidth]{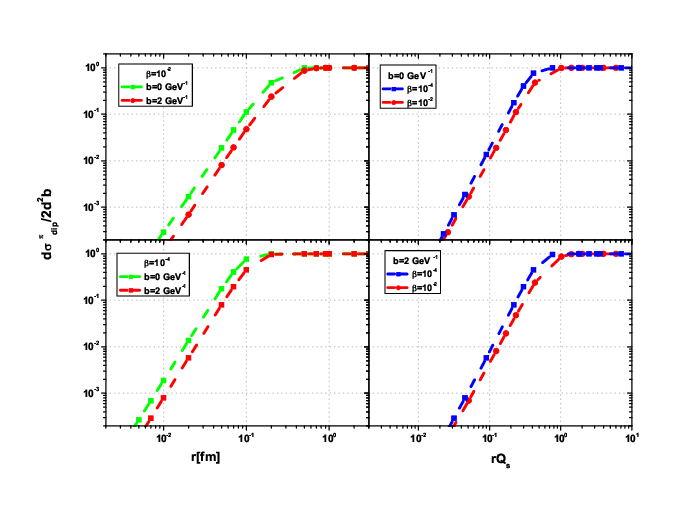}
\caption{Dipole-pion cross section in the bSat model as a function
of $r$ (left panel) and $rQ_{s}$ (right panel) for $\beta=10^{-2}$
and $10^{-4}$ at the impact parameters $b=0$ and
$2~\mathrm{GeV}^{-1}$.
 Left panel (upper): $\beta=10^{-2}$ and
$b=0~\mathrm{GeV}^{-1}$ (green-square) and $b=2~\mathrm{GeV}^{-1}$
(red-circle).
 Left panel (lower): $\beta=10^{-4}$ and
$b=0~\mathrm{GeV}^{-1}$ (green-square) and $b=2~\mathrm{GeV}^{-1}$
(red-circle). Right panel (upper): $b=0~\mathrm{GeV}^{-1}$ and
$\beta=10^{-4}$ (blue-square) and $\beta=10^{-2}$ (red-circle).
Right panel (lower): $b=2~\mathrm{GeV}^{-1}$ and $\beta=10^{-4}$
(blue-square) and $\beta=10^{-2}$ (red-circle).}\label{Fig1}
\end{figure}
\begin{figure}
\includegraphics[width=0.8\textwidth]{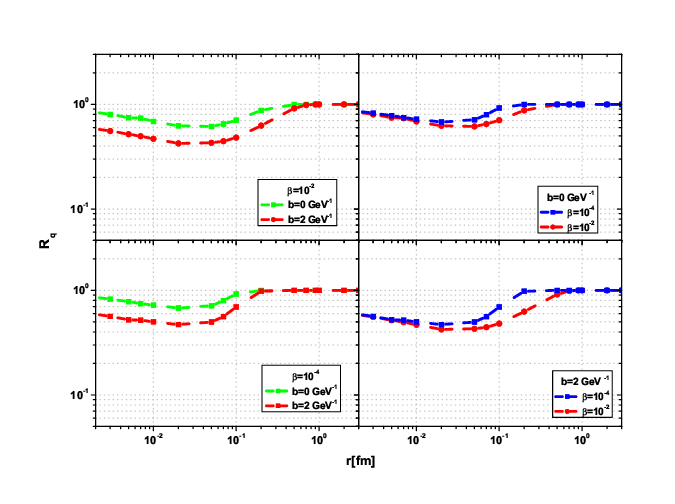}
\caption{The ratio $
R_{q}={\frac{d\sigma^{\pi}_{\mathrm{dip}}}{d^{2}b}(\mathbf{b},\mathbf{r},\beta)}/{\frac{d\sigma^{p}_{\mathrm{dip}}}{d^{2}b}(\mathbf{b},\mathbf{r},\beta)}
$ is plotted as a function of $r$ for $\beta=10^{-2}$ and
$10^{-4}$ at the impact parameters $b=0$ and
$2~\mathrm{GeV}^{-1}$.
 Left panel (upper): $\beta=10^{-2}$ and
$b=0~\mathrm{GeV}^{-1}$ (green-square) and $b=2~\mathrm{GeV}^{-1}$
(red-circle).
 Left panel (lower): $\beta=10^{-4}$ and
$b=0~\mathrm{GeV}^{-1}$ (green-square) and $b=2~\mathrm{GeV}^{-1}$
(red-circle). Right panel (upper): $b=0~\mathrm{GeV}^{-1}$ and
$\beta=10^{-4}$ (blue-square) and $\beta=10^{-2}$ (red-circle).
Right panel (lower): $b=2~\mathrm{GeV}^{-1}$ and $\beta=10^{-4}$
(blue-square) and $\beta=10^{-2}$ (red-circle).}\label{Fig2}
\end{figure}

In conclusion, we have discussed the evolution of the dipole-pion
cross section using the Laplace transform method in the bSat
model. This evolution is presented as a function of the dipole
size and the scaled Bjorken variable $\beta$. This method can
successfully describe the behavior of the dipole-pion cross
sections in terms of the impact parameter at small $\beta$ (Fig.1,
left). The scaling behavior in the bSat model is shown at large
and small dipole sizes, similar to the GBW model. These results in
$rQ_{s}$ (Fig.1, right), which are independent of the impact
parameter and the scaled Bjorken variable, show that the evolution
of DGLAP based on the Laplace transform into the initial scale
plays a dominant role in the dipole-pion cross sections, as the
results
show a geometrical scaling in a wide range of $rQ_{s}$.\\
The ratio $R_{q}$ in the bSat model  depends on the impact
parameter at small dipole sizes and is scaled at large dipole
sizes (Fig.2). This ratio varies between the values
$0.4<R_{q}{\leq}1$ over a wide range of $r$. The behavior of the
ratio $R_{q}$ over this range of $r$  depends on the structure of
the proton, from the valence quarks to a multiquark in the proton.
Quantum fluctuations in the proton at small dipole sizes interact
with the quarks and antiquarks in the meson cloud of the proton,
depending on the flavor asymmetric and flavor symmetric sea
\cite{Kopeliovich}.\\

\subsection{ACKNOWLEDGMENTS}

The work of B.Kopeliovich was supported in part by Grant ANID -
Chile FONDECYT 1231062 , by ANID PIA/APOYO AFB230003.\\


\end{document}